# Effect of Coherence on Radiation Forces acting on a Rayleigh Dielectric Sphere

**Li-Gang Wang**

*Department of Physics, Zhejiang University, Hangzhou, 310027, China*

*and Department of Physics, Hong Kong Baptist University, Kowloon Tong, Hong Kong*

**Cheng-Liang Zhao, Li-Qin Wang, and Xuan-Hui Lu**

*Department of Physics, Zhejiang University, Hangzhou, 310027, China*

**Shi-Yao Zhu**

*Department of Physics, Hong Kong Baptist University, Kowloon Tong, Hong Kong*

The radiation forces on a Rayleigh dielectric sphere induced by a partially coherent light beam are greatly affected by the coherence of the light beam. The magnitude of the radiation forces on a dielectric sphere near the focus point greatly decreases as the coherence decreases. For the light beam with good coherence, the radiation force may be used to trap a particle; and for the light beam with intermediate coherence, the radiation force may be used to guide and accelerate a particle. © 2006 Optical Society of America.

Optical trapping and manipulation of micron-sized particles have been extensively attended (see some reviews [1-4]) since the seminal work of Ashkin [5] on radiation pressure. This technology has been a powerful tool, in particular, in biophysical sciences for trapping living cells and organelles [6-7], DNA analysis [8], and atomic physics for manipulating the neutral atoms [9-10].

In 1970, Ashkin [5] first demonstrated that micron-sized particles can be accelerated and trapped by the radiation forces from two counter-propagating laser beams. In 1986, Ashkin et al. further showed that even a single laser beam focused into the dielectric sphere can pull up and trap this dielectric sphere at the focus point [11]. The trap induced by a single beam is termed as a single-beam gradient force trap. Up to now, there are a series of investigation on the trapping characteristics of various of single-beam gradient force traps induced by different kinds of beams [5,11-18], such as Gaussian [5, 11], bottle [13], zero-order Bessel [15], and self-focused laser beams [16], and even evanescent fields [17-18]. In all these previous studies attention was paid on the fully coherent light, which can be highly focused or enhanced by different instruments such as the lens with high numerical aperture [5,11-12] and a sharply pointed metal tip [19]. However, in practice, any laser fields are always partially coherent [20]. It is well known that the focusing characteristics of partially coherent light beams are greatly affected by their coherence [21-23]. Therefore, it can be expected that radiation forces induced by partially coherent light beams will also depends on their coherence.

In this Letter we consider the effect of coherence of a highly focused Gaussian-Schell model (GSM) beam on the radiation force acting on a Rayleigh dielectric sphere. The profiles of the intensity and the degree of spatial coherence of the beam are Gaussian [20]. Many authors have pointed out that the intensity distributions and focusing characteristics for a GSM beam near the focus of a lens are greatly affected by its coherence [21, 24-25]. We ask ourselves what is the influence of the coherence of a GSM beam on the radiation force.

For simplicity, we consider a two-dimensional (2D) GSM beam (with the wavelength $\lambda_0$ in vacuum) focused by a thin lens with the focus length $f$, as shown in Fig. 1. From the theory of coherence, the cross-spectral density of the incident GSM beam at the input plane (z=0) can be assumed to be [20]:

$$W_{in}(x_1, x_2, z=0) = I_0 \exp\left(-\frac{x_1^2 + x_2^2}{w_0^2}\right) \exp\left(-\frac{(x_1 - x_2)^2}{2\sigma_0^2}\right). \tag{1}$$

Here $x_1$ and $x_2$ denote the coordinates of two typical points on the x axis ($z = 0$), $w_0$ is the spot size, and $\sigma_0$ is the correlation width representing the spatial coherence of the GSM beam, $I_0$ is a constant. From Eq. (1), the intensity is $I(x, z = 0) = W_{in}(x_1 = x, x_2 = x, z = 0) = I_0 \exp\left(-\frac{2x^2}{w_0^2}\right)$, which is independent of coherence, thus such beams have the same intensity profiles with different coherence at $z = 0$. By using the generalized Huygens-Fresnel diffraction integral [24], the intensity distribution of the GSM beam passing through a thin lens (see in Fig. 1) can be expressed as

$$I_{out}(x, z) = \frac{I_0}{\Theta^{1/2}} \exp\left(-\frac{2x^2}{w_0^2 \Theta}\right), \tag{2}$$

where $\Theta = \Delta z^2 / f^2 + (z/z_0)^2 (1 + w_0^2/\sigma_0^2)$, $\Delta z = z - f$ is the distance between the geometrical focusing point $F$ and output plane, $z_0 = \pi w_0^2 / \lambda$ is the Rayleigh distance for the coherent Gaussian beam in the surrounding medium with the refractive index $n_m$, and $\lambda = \lambda_0 / n_m$ is the wavelength in the surrounding medium. From Eq. (2), it is clear that although the beams with different coherence have the same intensity profiles in the plane of $z = 0$ (i.e., at the lens plane), the output intensity distribution after the lens will be greatly affected by the correlation width $\sigma_0$ (i.e., its coherence).

It is well known that light has the momentum, and the radiation force comes from the momentum transfer between photons and particles. There are two types of radiation forces: the scattering and gradient forces [4, 11]. For single-beam gradient traps both two forces will be configured to give a point of stable equilibrium located close to the beam focus [11], which can be used to trap a particle. In this paper, we assume the diameter $2a$ of a dielectric particle (with refractive index $n_p$) be much smaller than $\lambda_0$, so that the scattering and gradient forces can be expressed, respectively, as [4,26]: $\vec{F}_{Scat} = \vec{e}_z \frac{n_m}{c} \alpha I_{out}$ and $\vec{F}_{Grad} = \frac{2\pi n_m}{c} \beta \nabla I_{out}$, where $\alpha = \frac{128\pi^5 a^6}{3\lambda^4} \left(\frac{m^2 - 1}{m^2 + 2}\right)^2$ with $m = n_p / n_m$ is the scattering cross section of a spherical particle, $\beta = a^3 \left(\frac{m^2 - 1}{m^2 + 2}\right)$ is the polarizability of a spherical particle. Obviously, the scattering force is along the propagating direction of the light beam, and could be directly given in term of the output intensity distribution by

$$\vec{F}_{Scat} = \vec{e}_z \frac{n_m \alpha I_0}{c \Theta^{1/2}} \exp\left(-\frac{2x^2}{w_0^2 \Theta}\right). \tag{3}$$

From the above, we can also obtain the transverse and longitudinal gradient forces as follows:

$$\vec{F}_{Grad,x} = -\vec{e}_x \frac{8\pi n_m \beta I_0 x}{c w_0^2 \Theta^{3/2}} \exp\left(-\frac{2x^2}{w_0^2 \Theta}\right), \tag{4a}$$

$$\vec{F}_{Grad,z} = -\vec{e}_z \frac{2\pi n_m \beta}{c} \frac{I_0 \left[\frac{\Delta z}{f^2} + \frac{z}{z_0^2}(1 + \frac{w_0^2}{\sigma_0^2})\right]\left[\Theta - \frac{2x^2}{w_0^2}\right]}{\Theta^{5/2}} \exp\left(-\frac{2x^2}{w_0^2 \Theta}\right). \tag{4b}$$

Obviously, for the 2D light beam, the gradient force consists of two components that act as restoring forces directed towards the real focusing center of the beam for the particle with $m > 1$. From Eqs. (3), (4a-b), both the scattering and gradient forces will be greatly affected by the coherence of the beam.

In the following calculations, we take the parameters as that in Ref. [26]: $\lambda_0 = 0.5415$ μm, $n_m = 1.332$ (water) and $n_p = 1.592$; and the intensity $I_0$ of the incident light (with $w_0 = 2$ mm) is assumed to be $I_0 = 200$ mW/μm$^2$. The focusing length of the lens is $f = 2$ cm. Figure 2 shows the typical effect of the coherence on both the scattering and gradient forces acting on the Rayleigh particle with radius $a = 10$ nm. From Fig. 2 (a), (b) and (c), for the fully coherent beams, the magnitude of the radiation forces on the dielectric particles are the strongest (e. g., see the solid curves for $\sigma_0 = 5w_0$), in this case, the radiation force may be used as optical trapping and manipulate the particle with $m > 1$. The magnitude of the radiation forces decrease quickly with $\sigma_0$ becoming smaller [see the red dashed curves for $\sigma_0 = 1.0w_0$, the blue dash-dotted curves for $\sigma_0 = 0.5w_0$, and the green short-dashed curves for $\sigma_0 = 0.2w_0$ in Fig. 2 (a-c)]. Figure 2(d) shows that there is a stable equilibrium near the focusing point for the case of $\sigma_0 = 5w_0$. With the decreasing of $\sigma_0$, the stable equilibrium region of the radiation force near the focusing point gradually disappears [see the curve in (e) $\sigma_0 = w_0$, (f) $\sigma_0 = 0.5w_0$ and (g) $\sigma_0 = 0.2w_0$ of Fig. 2]. In Fig. 2(g), the scattering force is dominated in the z direction so that the particle may be accelerated along the beam propagation. Therefore, one can control the radiation forces by adjusting the light coherence of the incident beams with the same intensities.

In practice, the particle in the surrounding medium (such as in water) always suffers the Brownian motion due to the thermal fluctuations. From the fluctuation-dissipation theorem of Einstein, the magnitude of the Brownian force $\vec{f}_B$ is given by [17] $\sqrt{<\vec{f}_B \cdot \vec{f}_B>_t} = (12\pi\eta a k_B T)^{1/2}$, where the subscript $t$ denotes the time average, $\eta$ is the viscosity of water ($\eta = 7.977 \times 10^{-4} Pa \cdot s$ at room temperature of $30°C$ in our calculation), $k_B$ is the Boltzmann constant, and $T$ is the temperature of water. In order to trap the particle stably, we have to compare the magnitudes of the maximum radiation force and the Brownian force. From Eq. (2), we can readily find the real focusing point $F'$ should be at $F' = (0, f/[1 + f^2(1 + w_0^2/\sigma_0^2)/z_0^2])$ instead of the geometric focusing point $F = (0, f)$. In the real focusing plane at $z = f/[1 + f^2(1 + w_0^2/\sigma_0^2)/z_0^2]$, the maximum transverse gradient force can be expressed by

$$\left|F_{Grad,x}\right|_{A,B}^{max} = \frac{4\pi n_m \beta I_0 \exp[-1/2]}{cw_0}\left[1 + \frac{z_0^2}{f^2}\frac{\sigma_0^2}{\sigma_0^2 + w_0^2}\right] \quad (7)$$

at two points $A, B = (\pm w_0\{1 + z_0^2\sigma_0^2/[f^2(\sigma_0^2 + w_0^2)]\}^{-1/2}/2, f/[1 + f^2(1 + w_0^2/\sigma_0^2)/z_0^2])$. Along the z axis, the longitudinal gradient force reaches its maximum

$$\left|F_{Grad,z}\right|_{C,D}^{max} = \frac{4\pi n_m \beta}{c} \cdot \frac{I_0\left[z_0^2 + f^2(1 + w_0^2/\sigma_0^2)\right]^{3/2}}{3\sqrt{3}z_0 f^3(1 + w_0^2/\sigma_0^2)} \quad (8)$$

at two points $C, D = (0, f[1 \pm f(1 + w_0^2/\sigma_0^2)^{1/2}/(\sqrt{2}z_0)]/[1 + f^2(1 + w_0^2/\sigma_0^2)/z_0^2])$, where the scattering forces are given by $\left|\vec{F}_{Scat}\right|_{C,D} = \vec{e}_z \frac{n_m \sigma}{c}\sqrt{2/3}I_0\{1 + z_0^2\sigma_0^2/[f^2(\sigma_0^2 + w_0^2)]\}^{1/2}$. The necessary criterion for axial stability of a single-beam trap is that $R$, the ratio of the backward gradient force to the forward-scattering force, is greater than unity at the position of the maximum longitudinal gradient force [11], i.e., $R = \left|F_{Grad,z}\right|_{C,D}^{max}/\left|\vec{F}_{Scat}\right|_{C,D} \geq 1$. Obviously, this condition is also dependent on $\sigma_0$. By comparing

the radiation forces at these critical points (A, B, C and D) with the Brownian force, one can judge whether the particle can be trapped or not.

Figure 3 shows the dependence of the maximum transverse and longitudinal gradient forces on $\sigma_0$ with different radius $a$. For the comparison, we also plot the Brownian force at temperature $30°C$ and the scattering force at the position where the longitudinal gradient force on the z axis reaches its maximum. Both the maximum transverse and longitudinal gradient forces decrease as the coherence decreases [see the solid and dashed curves in Fig. 3 (a)-(d)], and the scattering force is also decreasing with the reduction of $\sigma_0$ [see the dot-dashed curves in Fig. 3 (a)-(d)]. For the different-size particles, the magnitude of the Brownian force is proportional to $a^{1/2}$ [see the dotted lines in Fig. 3(a)-(d)]. When the particle is very small [see Fig. 3(a)], for the beams with the large value of $\sigma_0$ (good coherence), the radiation forces are dominated by $\vec{F}_{Grad,x}$ and $\vec{F}_{Grad,z}$ [see the right side of the vertical line $P$ in Fig. 3(a)], so the radiation forces can stably trap and manipulate the small particle in this case. For the beams with the intermediate value of $\sigma_0$ between the vertical lines $P$ and $Q$ in Fig. 3(a) (where $\vec{F}_{Grad,x}$ can overcome $\vec{f}_B$), if $\vec{F}_{Scat}$ can also overcome $\vec{f}_B$ [also see Fig. 3(b)], the particle will be guided and accelerated along the beam propagation, conversely, the particle will escape away from the beam trapping even if it can be confined in the transverse direction due to the transverse gradient force. For the beams with low coherence [see the left side of the line $Q$ in Fig. (a)-(d)], the radiation force cannot be used to trap or accelerate any kind of particles due to the effect of the Brownian motion. For the large particles (in the Rayleigh approximation), $\vec{F}_{Scat}$ will be larger than $\vec{F}_{Grad,z}$, see Fig. 3(c) and (d), in this case, the radiation force produced by the beam with good coherence can be easily used to accelerate the particle along the beam propagation.

In conclusion, we have investigated the radiation forces on the Rayleigh dielectric particle produced by the partially coherent light beam. It is shown that the radiation forces are greatly affected by the coherence of the light. As the coherence decreases, the magnitude of the radiation forces on a dielectric sphere near the focus point greatly decreases. Our calculations make clear that whether the light beam can be used to trap the particle is greatly affected by the beam coherence. For the light beam with good coherence, the radiation force may be used to trap a particle; for the light beam with intermediate coherence, the radiation force could be used to guide a particle along the beam propagation, while for the low-coherent light beam, it is very difficult to manipulate a particle.

This work was supported by Scientific Research Foundation for the Returned Overseas Chinese Scholars of Zhejiang Province (G80611), and partially supported by National Natural Science Foundation of China (under contract No. 10547138 and 10604047) and RGC (HKBU2027/04P). Li-Gang Wang's e-mail address is sxwlg@yahoo.com.cn.

# Figure Captions

FIG. 1. Schematic of a GSM beam focused onto a particle (with index $n_p$) within the surrounding medium (with index $n_m$). Point $F$ is the geometrical focus point, and the lens is located at the input plane.

FIG. 2. (Color Online) The Effect of coherence on (a) $F_{Grad,x}$, (b) $F_{Grad,z}$ and (c) $F_{Scat}$. In (a-c), the red dashed curves, $\sigma_0 = 1.0 w_0$, the blue dash-dotted curves, $\sigma_0 = 0.5 w_0$, and the green short-dashed curves, $\sigma_0 = 0.2 w_0$. The changes of the radiation force (the sum of $\vec{F}_{Scat}$ and $\vec{F}_{Grad,z}$) along the $z$ axis for (d) $\sigma_0 = 5 w_0$, (e) $\sigma_0 = 1.0 w_0$, (f) $\sigma_0 = 0.5 w_0$ and (g) $\sigma_0 = 0.2 w_0$.

FIG. 3. The dependence of the values of $\left|\vec{F}_{Grad,x}\right|_{A,B}^{max}$ (solid), $\left|\vec{F}_{Grad,z}\right|_{C,D}^{max}$ (dashed) and $\left|\vec{F}_{Scat}\right|_{C,D}$ (dot-dashed) on the correlation width $\sigma_0$. Horizontal dotted lines denote the magnitudes of the Brownian forces. Vertical dotted lines P in (a) and (b) pass through the point of $\left|\vec{F}_{Grad,z}\right|_{C,D}^{max} = \left|\vec{F}_{Scat}\right|_{C,D}$ and vertical dotted lines Q in (a-d) go through the cross points of the solid curves and horizontal dotted lines.

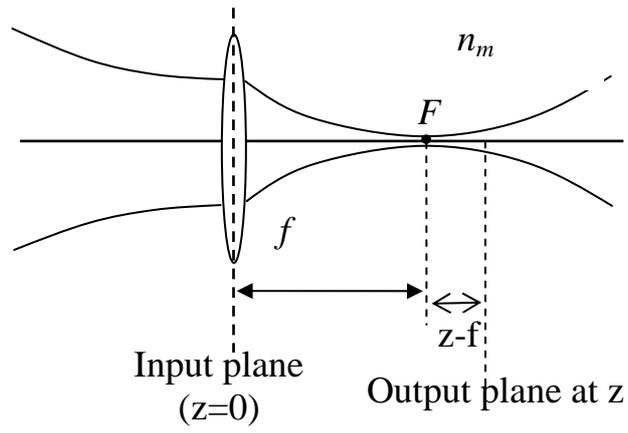

**FIG. 1**

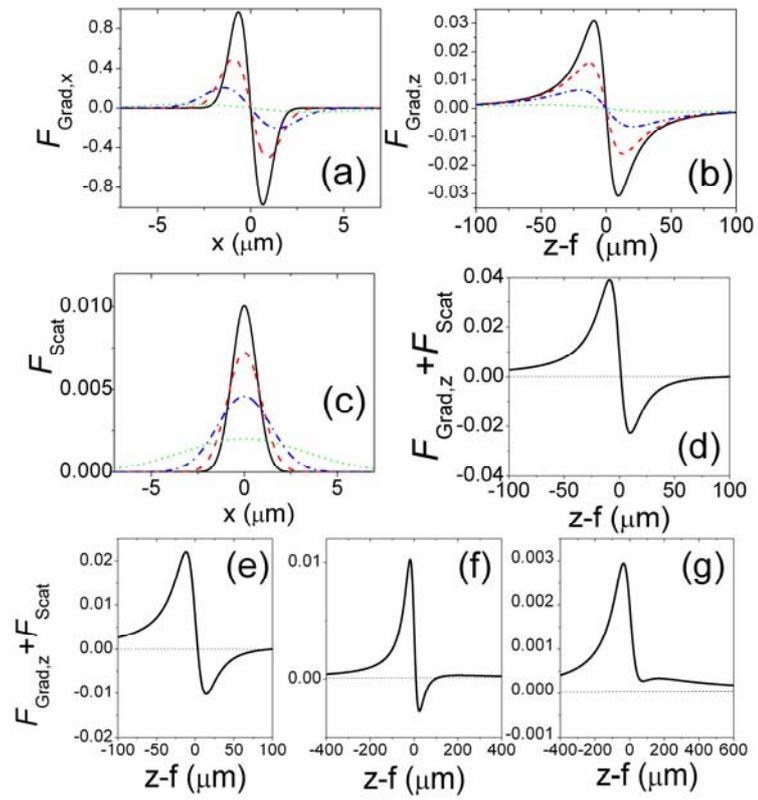

**FIG. 2**

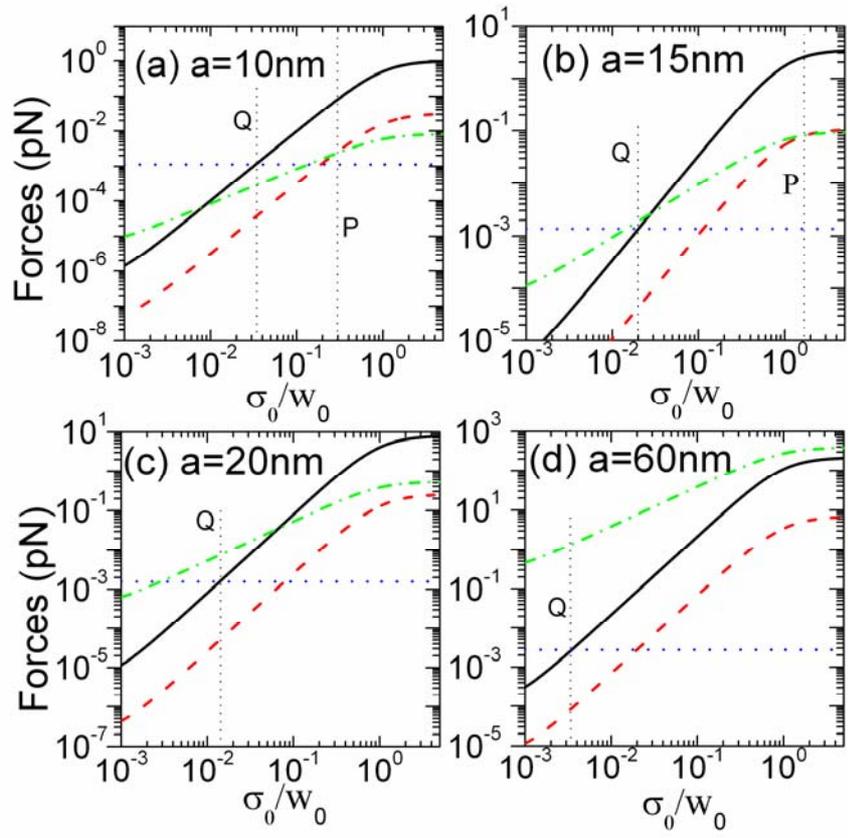

**FIG. 3**